\documentclass[showpacs,preprintnumbers,amsmath,prl,graphicx,amssymb,superscriptaddress,twocolumn]{revtex4}
\usepackage{graphicx}
\usepackage{dcolumn}
\usepackage{bm}
\usepackage{psfrag}
\usepackage{txfonts}
\newcommand{\snoccfluxunc}{1.67^{+0.05}_{-0.04}\mbox{(stat)}^{+0.07}_{-0.08}\mbox{(syst)}} 
\newcommand{\snoesfluxunc}{1.77^{+0.24}_{-0.21}\mbox{(stat)}^{+0.09}_{-0.10}~\mbox{(syst)}} 
\newcommand{\snoncfluxunc}{5.54^{+0.33}_{-0.31}~\mbox{(stat)}^{+0.36}_{-0.34}~\mbox{(syst)}} 
\newcommand{\snoccncratio}{0.301\pm0.033~\mbox{(total)}}
\newcommand{\nccfit}{1867^{+91}_{-101}} 
\newcommand{\nesfit}{171^{+24}_{-22}} 
\newcommand{\nncfit}{983^{+77}_{-76}}  
\newcommand{\nncpfit}{267^{+24}_{-22}}
\newcommand{\nncbefit}{185^{+25}_{-22}}
\newcommand{\nncpbefit}{77^{+12}_{-10}}
\newcommand{\nncalphafit}{6127\pm101}

\renewcommand{\today}{\number\day\space\ifcase\month\or January\or 
 February\or March\or April\or May\or June\or July\or August\or 
 September\or October\or November\or December\fi\space\number\year}

\begin{document}
\title{Independent Measurement of the Total Active $^{\bm{8}}$B Solar Neutrino Flux Using an Array of $^{\bm{3}}$He Proportional Counters at the Sudbury Neutrino Observatory}
%
\newcommand{\alta}{Department of Physics, University of 
Alberta, Edmonton, AB, T6G 2R3, Canada}
\newcommand{\ubc}{Department of Physics and Astronomy, University of 
British Columbia, Vancouver, BC V6T 1Z1, Canada}
\newcommand{\bnl}{Chemistry Department, Brookhaven National 
Laboratory,  Upton, New York 11973-5000, USA}
\newcommand{\carleton}{Ottawa-Carleton Institute for Physics, Department of Physics, Carleton University, Ottawa, ON K1S 5B6, Canada}
\newcommand{\uog}{Physics Department, University of Guelph,  
Guelph, ON N1G 2W1, Canada}
\newcommand{\lu}{Department of Physics and Astronomy, Laurentian 
University, Sudbury, ON P3E 2C6, Canada}
\newcommand{\lbnl}{Institute for Nuclear and Particle Astrophysics and 
Nuclear Science Division, Lawrence Berkeley National Laboratory, Berkeley, California 94720, USA}
\newcommand{\lbla}{ Lawrence Berkeley National Laboratory, Berkeley, California, USA}
\newcommand{\lanl}{Los Alamos National Laboratory, Los Alamos, New Mexico 87545, USA}
\newcommand{\llnl}{Lawrence Livermore National Laboratory, Livermore, CA}
\newcommand{\lanla}{Los Alamos National Laboratory, Los Alamos, New Mexico 87545, USA}
\newcommand{\oxford}{Department of Physics, University of Oxford, 
Denys Wilkinson Building, Keble Road, Oxford OX1 3RH, United Kingdom}
\newcommand{\penn}{Department of Physics and Astronomy, University of 
Pennsylvania, Philadelphia, Pennsylvania 19104-6396, USA}
\newcommand{\queens}{Department of Physics, Queen's University, 
Kingston, ON K7L 3N6, Canada}
\newcommand{\uw}{Center for Experimental Nuclear Physics and Astrophysics, 
and Department of Physics, University of Washington, Seattle, Washington 98195, USA}
\newcommand{\uta}{Department of Physics, University of Texas at Austin, Austin, Texas 78712-0264, USA}
\newcommand{\triumf}{TRIUMF, 4004 Wesbrook Mall, Vancouver, BC V6T 2A3, Canada}
\newcommand{\ralimp}{Rutherford Appleton Laboratory, Chilton, Didcot OX11 0QX, UK}
\newcommand{\iusb}{Department of Physics and Astronomy, Indiana University, South Bend, IN}
\newcommand{\fnal}{Fermilab, Batavia, IL}
\newcommand{\uo}{Department of Physics and Astronomy, University of Oregon, Eugene, OR}
\newcommand{\hu}{Graduate School of Engineering, Hiroshima University, Hiroshima, Japan}
\newcommand{\slac}{Stanford Linear Accelerator Center, Menlo Park, CA}
\newcommand{\mac}{Department of Physics, McMaster University, Hamilton, ON}
\newcommand{\doe}{US Department of Energy, Germantown, MD}
\newcommand{\lund}{Department of Physics, Lund University, Lund, Sweden}
\newcommand{\mpi}{Max-Planck-Institut for Nuclear Physics, Heidelberg, Germany}
\newcommand{\uom}{Ren\'{e} J.A. L\'{e}vesque Laboratory, Universit\'{e} de Montr\'{e}al, Montreal, PQ}
\newcommand{\cwru}{Department of Physics, Case Western Reserve University, Cleveland, OH}
\newcommand{\pnnl}{Pacific Northwest National Laboratory, Richland, WA}
\newcommand{\uc}{Department of Physics, University of Chicago, Chicago, IL}
\newcommand{\mitt}{Laboratory for Nuclear Science, Massachusetts Institute of Technology, Cambridge, Massachusetts 02139, USA}
\newcommand{\ucsd}{Department of Physics, University of California at San Diego, La Jolla, CA }
\newcommand{	\lsu	}{Department of Physics and Astronomy, Louisiana State University, Baton Rouge, Louisiana 70803, USA}
\newcommand{\imp}{Imperial College, London SW7 2AZ, UK}
\newcommand{\uci}{Department of Physics, University of California, Irvine, CA 92717}
\newcommand{\ucia}{Department of Physics, University of California, Irvine, CA}
\newcommand{\suss}{Department of Physics and Astronomy, University of Sussex, Brighton  BN1 9QH, UK}
\newcommand{	\lifep	}{Laborat\'{o}rio de Instrumenta\c{c}\~{a}o e F\'{\i}sica Experimental de
Part\'{\i}culas, Av. Elias Garcia 14, 1$^{\circ}$, 1000-149 Lisboa, Portugal}
\newcommand{\hku}{Department of Physics, The University of Hong Kong, Hong Kong.}
\newcommand{\aecl}{Atomic Energy of Canada, Limited, Chalk River Laboratories, Chalk River, ON K0J 1J0, Canada}
\newcommand{\nrc}{National Research Council of Canada, Ottawa, ON K1A 0R6, Canada}
\newcommand{\princeton}{Department of Physics, Princeton University, Princeton, NJ 08544}
\newcommand{\birkbeck}{Birkbeck College, University of London, Malet Road, London WC1E 7HX, UK}
\newcommand{\snoi}{SNOLAB, Sudbury, ON P3Y 1M3, Canada}
\newcommand{\uba}{University of Buenas Aires, Argentina}
\newcommand{\hvd}{Department of Physics, Harvard University, Cambridge, MA}
\newcommand{\pny}{Goldman Sachs, 85 Broad Street, New York, NY}
\newcommand{\pnv}{Remote Sensing Lab, PO Box 98521, Las Vegas, NV 89193}
\newcommand{\psis}{Paul Schiffer Institute, Villigen, Switzerland}
\newcommand{\liverpool}{Department of Physics, University of Liverpool, Liverpool, UK}
\newcommand{\uto}{Department of Physics, University of Toronto, Toronto, ON, Canada}
\newcommand{\uwisc}{Department of Physics, University of Wisconsin, Madison, WI}
\newcommand{\psu}{Department of Physics, Pennsylvania State University,
     University Park, PA}
\newcommand{\anl}{Deparment of Mathematics and Computer Science, Argonne
     National Laboratory, Lemont, IL}
\newcommand{\cornell}{Department of Physics, Cornell University, Ithaca, NY}
\newcommand{\tufts}{Department of Physics and Astronomy, Tufts University, Medford, MA}
\newcommand{\ucd}{Department of Physics, University of California, Davis, CA}
\newcommand{\unc}{Department of Physics, University of North Carolina, Chapel Hill, NC}
\newcommand{\dresden}{Institut f\"{u}r Kern- und Teilchenphysik, Technische Universit\"{a}t Dresden,  01069 Dresden, Germany}
\newcommand{\isu}{Department of Physics, Idaho State University, Pocatello, ID}


\affiliation{\alta}
\affiliation{\ubc}
\affiliation{\bnl}
\affiliation{\carleton}
\affiliation{\uog}
\affiliation{\lu}
\affiliation{\lbnl}
\affiliation{\lifep}
\affiliation{\lanl}
\affiliation{\lsu}
\affiliation{\mitt}
\affiliation{\oxford}
\affiliation{\penn}
\affiliation{\queens}
\affiliation{\snoi}
\affiliation{\uta}
\affiliation{\triumf}
\affiliation{\uw}

\author{B.~Aharmim}\affiliation{\lu}
\author{S.N.~Ahmed}\affiliation{\queens}
\author{J.F.~Amsbaugh}\affiliation{\uw}
\author{A.E.~Anthony}\affiliation{\uta}
\author{J.~Banar}\affiliation{\lanl}
\author{N.~Barros}\affiliation{\lifep}
\author{E.W.~Beier}\affiliation{\penn}
\author{A.~Bellerive}\affiliation{\carleton}
\author{B.~Beltran}\affiliation{\alta}\affiliation{\queens}
\author{M.~Bergevin}\affiliation{\lbnl}\affiliation{\uog}
% ...
			
\collaboration{SNO Collaboration}
\noaffiliation

\date{June 5 2008}
\begin{abstract}
The Sudbury Neutrino Observatory (SNO) used an array of $^{3}$He proportional counters to measure the rate of neutral-current interactions in heavy water and precisely determined the total active ($\nu_{x}$) $^{8}$B solar neutrino flux.  This technique is independent of previous methods employed by SNO.  The total flux is found to be $\snoncfluxunc\times10^{6}$~cm$^{-2}$s$^{-1}$, in agreement with previous measurements and standard solar models.  A global analysis of solar and reactor neutrino results yields $\Delta m^{2} = 7.59^{+0.19}_{-0.21}\times10^{-5}$~eV$^2$ and $\theta = 34.4^{+1.3}_{-1.2}$ degrees.  The uncertainty on the mixing angle has been reduced from SNO's previous results.
\end{abstract}

\maketitle
The Sudbury Neutrino Observatory~\cite{sno_nim} detects $^8$B solar neutrinos
through three reactions:  charged-current interactions (CC) on deuterons, in which only electron neutrinos participate; neutrino-electron elastic scattering (ES), which are dominated by contributions from electron neutrinos; and neutral-current (NC) disintegration of the deuteron by neutrinos, which has equal sensitivity to all active neutrino flavors.

In its first phase of operation, SNO measured the NC rate by observing neutron captures on deuterons and found
that a Standard-Electro-Weak-Model description with an undistorted ${}^{8}$B neutrino spectrum
and CC, NC, and ES rates due solely to $\nu_{e}$ interactions
was rejected at $5.3\sigma$~\cite{ccprl,ncprl,dnprl,long}.    The second phase of SNO measured the rates and spectra after the addition of $\sim$2000 kg of NaCl to the $10^6$ kg of heavy water (D$_2$O).  This enhanced the neutron detection efficiency and the ability to statistically
separate the NC and CC signals, and resulted in significant improvement
in the accuracy of the measured  $\nu_e$ and $\nu_x$  fluxes without any assumption about the energy dependence of the neutrino flavor transformation~\cite{saltprl,nsp}.  In the present measurement, the NC signal neutrons were predominantly detected by an array of $^{3}$He proportional counters (Neutral Current Detection, or NCD, array~\cite{ncdnim}) consisting of 36 ``strings'' of counters that were deployed in the D$_2$O.  Four additional strings filled with $^{4}$He were insensitive to the neutron signals and were used to study backgrounds.  Cherenkov light signals from CC, NC, and ES reactions were still recorded by the photomultiplier tube (PMT) array, though the rate of such NC events from $^2$H(n,$\gamma$)$^3$H reactions was significantly suppressed due to neutron absorption in the $^{3}$He strings.  As described in this Letter, the new measurements of the CC, NC, and ES rates result in reduced correlation between the fluxes and improvement in the mixing angle uncertainty.

The data presented here were recorded between November 27, 2004 and November 28, 2006, totaling 385.17 live days.  The number of raw NCD triggers was 1,417,811 and the data set was  reduced to 91,636 NCD events after data reduction described in~\cite{ncdnim}.  Six strings filled with $^{3}$He were excluded from the analysis due to various defects.  The number of raw PMT triggers was 146,431,347 with 2381 PMT events passing data reduction and analysis selection requirements similar to those in~\cite{long}.  Background events arising from $\beta$--$\gamma$ decays were reduced by selecting events with reconstructed electron effective kinetic energies $\geq 6.0$~MeV and reconstructed vertices within $R_{\rm{fit}} \leq 550$~cm.

Thermal neutron capture on the $^{3}$He in the proportional counters results in the creation of a proton-triton pair with a total kinetic energy of 764 keV.  Because of particles hitting the counter walls~\cite{ncdnim},  the detected ionization energy was between 191 and 764 keV.  The signals from each string were amplified logarithmically to provide sufficient dynamic range before they were digitized~\cite{ncdnim}.  The detectors were constructed from ultra-pure nickel produced by a chemical vapor deposition process to minimize internal radioactivity.

The neutron detection efficiency and response of the PMT and NCD arrays have been determined with a variety of neutron calibration sources.  Neutron point sources (${}^{252}$Cf and ${}^{241}$AmBe) were frequently deployed throughout the detector volume to measure the temporal stability and the detector gain of the NCD array.  The NC neutron detection efficiency was studied by using an isotropic source of neutrons produced by mixing $^{24}$Na (t$_{1/2}$ = 14.959 hours), in the form of activated NaCl, into the heavy water in October 2005 and October 2006.  Neutrons were produced by deuterium photodisintegration induced by the 2.754-MeV $^{24}$Na gammas.  The largest uncertainties on the neutron detection efficiency were associated with the knowledge of the $^{24}$Na source strength and the ability to determine the uniformity of its mixing in the heavy water.  The inferred NC neutron capture efficiency for the NCD array was 0.211$\pm$0.007 in good agreement with the 0.210 $\pm$ 0.003 given by a Monte Carlo simulation verified against point-source data.  The fraction of detected neutrons inside the analysis energy range from 0.4--1.4~MeV, including the effects of data reduction, electronic thresholds and efficiency, and digitizer live time, was 0.862$\pm$0.004.  The neutron detection efficiency for the PMT array was 0.0485$\pm$0.0006 determined from neutron point sources.

\begin{figure}
\begin{center}
\includegraphics[width=3.2in]{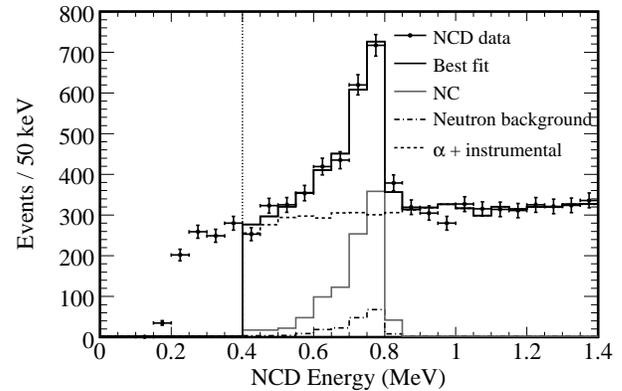}
\caption{\label{ncd_energy} NCD energy spectrum fit with a neutron calibration spectrum, neutron backgrounds, alpha background derived from Monte Carlo simulation, and low-energy instrumental background distributions.  Data are shown after data reduction up to 1.4~MeV, and the fit is above 0.4~MeV.}
\end{center}
\vspace{-4ex}
\end{figure}

The energy spectrum of the reduced NCD data set is shown in Fig.~\ref{ncd_energy}.  The distinctive neutron spectrum peaks at 764 keV.  This spectrum was fit with a neutron energy spectrum taken from the $^{24}$Na calibration.  The alpha background distribution was derived from a Monte Carlo simulation of the proportional counters.  The alpha background energy spectrum has several components, U and Th progeny in the bulk of the nickel detector bodies and $^{210}$Po on the inner surfaces~\cite{ncdnim}.  These sources resulted in approximately 16 alphas per day detected in the full neutron energy window for the entire NCD array. The Monte Carlo simulation was verified using alpha data from the array above 1.2 MeV and from the $^4$He strings in the neutron energy region.  Several uncertainties were included in the alpha background distribution: depth profile and composition of alpha emitters in nickel, electron drift time, space-charge model parameters, and ion mobility.  Low-energy instrumental background events were found on two strings that were excluded from the analysis.  Distributions of these events were used to fit for possible additional contamination in the data on the rest of the array.  

The optical and energy responses and position and directional reconstruction of the PMT array were updated to include light shadowing and reflection from the NCD array.  With improvements to the calibration data analysis and increased high voltage on the PMT array, the introduction of the NCD array did not significantly increase the position or energy reconstruction uncertainties from previous phases.  A  normalization for the photon detection efficiency based on ${}^{16}$N calibration data~\cite{n16_nim} and Monte Carlo simulations was used to set the absolute energy scale.  The energy response for electrons can be characterized by a Gaussian function with resolution $\sigma_{T} = -0.2955 + 0.5031\sqrt{{T}_{e}}+0.0228\hspace{0.03in}{T}_{e}$, where ${T}_{e}$ is the electron kinetic energy in MeV.  The energy scale uncertainty was $1.1\%$.

Backgrounds are summarized in Table~\ref{letable}.  Low levels of ${}^{214}$Bi and ${}^{208}$Tl present in the heavy and light water, NCD counters, and cables can create free neutrons from deuteron photodisintegration and low-energy Cherenkov events from $\beta-\gamma$ decays.  Techniques to determine these backgrounds in the water are described in previous works~\cite{ncprl,mnox_nim,htio_nim,rn_nim}.  In addition, alphas from Ra progeny on the NCD tube surfaces can induce $^{17,18}$O($\alpha$,n) interactions.  The background contributions from the NCD array were determined by combining the analyses of the alpha energy spectrum and the time-correlated alpha events in the decay chains.  The results from these studies agreed with those from radioassays of the materials prior to the construction of the NCD array.  In addition, \textit{in situ} analysis of the Cherenkov light found three detectable ``hotspots'' of elevated radioactivity on two strings.  Evaluations of their isotopic composition were made upon removal of the NCD array after the end of data taking.  
Results from the \textit{in situ} and \textit{ex situ} methods showed the neutron background uncertainty from the hotspots to be less than 0.7\% of the NC signal.  Neutron backgrounds from atmospheric neutrino interactions and ${}^{238}$U fission were estimated with NUANCE~\cite{casper} and from event multiplicities. 

Previous results~\cite{saltprl,nsp} reported the presence of external-source neutrons from the acrylic vessel and light water.  Alpha radioactivity measurements of  the acrylic vessel neck with Si counters before and after the NCD phase showed values consistent with those from the previous phase within uncertainties.  Thus, the external-source neutron contribution from the vessel was taken to be the same as for the previous phase.  The contributions from the light water were determined from the measured ${}^{214}$Bi and ${}^{208}$Tl concentrations. 

Backgrounds from Cherenkov events inside and outside the fiducial volume were estimated using calibration source data, measured activities, Monte Carlo calculations, and controlled injections of Rn into the detector.  These backgrounds were found to be small above the analysis energy threshold and within the fiducial volume, and were included as an additional uncertainty on the flux measurements.  Previous phases identified isotropic acrylic vessel background (IAVB) events, which can be limited to 0.3 remaining IAVB events (68\% CL) after data reduction for this phase. 

\begingroup
\begin{table}
\squeezetable
\caption{\label{letable}Background events for the PMT and NCD arrays, respectively. Backgrounds with similar detection efficiencies are listed together. The internal and external neutrons and the $\gamma$-ray backgrounds are constrained in the analysis. ``Other backgrounds'' include terrestrial $\bar{\nu}$s, reactor $\bar{\nu}$s, spontaneous fission, cosmogenics, CNO $\nu$s, and ($\alpha$,n) reactions. The last two entries are included in the systematic uncertainty estimates for the PMT array.}
\begin{ruledtabular}
\begin{tabular}{lll}
Source                                  & PMT Events    & NCD Events                    \\ \hline
D$_2$O radioactivity      &  $7.6 \pm 1.2$   &    $28.7 \pm 4.7$       \\
NCD bulk/$^{17}$O,$^{18}$O	& $4.6^{+2.1}_{-1.6}$ & $27.6^{+12.9}_{-10.3}$ \\
Atmospheric $\nu$/$^{16}$N        &  $ 24.7 \pm 4.6$   &       $ 13.6 \pm 2.7$        \\
``Other backgrounds''   &  $0.7 \pm 0.1$         &  $2.3 \pm 0.3$     \\
NCD hotspots		&	$17.7 \pm 1.8  $ 	&	$64.4 \pm 6.4$ 	\\  
NCD cables	& $1.1 \pm 1.0$ & $8.0 \pm 5.2$ \\ \hline
Total internal neutron background       &  $56.4^{+5.6}_{-5.4}$   &  $144.6^{+13.8}_{-14.8}$     \\ \hline 
External-source neutrons  &  $20.6 \pm 10.4$ & $40.9 \pm 20.6$ \\ \hline
Cherenkov events from $\beta$--$\gamma$ decays   & $5.8^{+9.7}_{-2.9}$ & \hspace{0.2in} ... \\ 
IAVB                                          & $<0.3$ (68\% CL) & \hspace{0.2in}      ... \\ 
\end{tabular}
\end{ruledtabular}
\end{table}
\endgroup

A blind analysis procedure was used to minimize the possibility of introducing biases.  The data set used during the development of the analysis procedures excluded a hidden fraction of the final data set and included an admixture of neutron events from muon interactions.  The blindness constraints were removed after all analysis procedures, parameters, and backgrounds were finalized. 
A simultaneous fit was made for the number of NC events detected by the NCDs, the numbers of NC, CC and ES events detected by the PMTs, as well as the numbers of background events of various types.	A Markov-chain Monte Carlo (MCMC) method with the Metropolis-Hastings algorithm~\cite{metropolis, hastings} was employed to make the fit, which also allowed nuisance parameters (systematics) weighted by external constraints to vary in the fit.  The NCD event energy spectrum was fit with an alpha background distribution, a neutron calibration spectrum, expected neutron backgrounds, and two instrumental background event distributions.  The PMT events were fit in reconstructed energy, the cosine of the event direction relative to the vector from the sun ($\cos\theta_{\odot}$), and the reconstructed radial position.  

The spectral distributions of the ES and CC events were not constrained to the ${}^{8}$B shape, but were extracted from the data.  Fits to the data yielded the following number of events: $\nncfit$ NC (NCD array), $\nncpfit$ NC (PMT array), $\nccfit$ CC, and $\nesfit$ ES, with $\nncbefit$ and $\nncpbefit$ neutron background events in the NCD and PMT arrays, respectively.  Additionally, the total NCD array background fits including alphas and the two instrumental components yielded $\nncalphafit$ events.

Assuming the $^8$B neutrino spectrum from~\cite{winter}, the equivalent neutrino fluxes derived from the fitted CC, ES, and NC events are (in units of $10^6~{\rm cm}^{-2} {\rm s}^{-1}$)~\cite{hep_footnote,es_crosssection}: 
\begin{eqnarray*}
\phi^{\text{SNO}}_{\text{CC}} & = & \snoccfluxunc \\
\phi^{\text{SNO}}_{\text{ES}} & = & \snoesfluxunc \\
\phi^{\text{SNO}}_{\text{NC}} & = & \snoncfluxunc~\mbox{,}
\end{eqnarray*} and the ratio of the ${}^{8}$B neutrino flux measured with
the CC and NC reactions is
\begin{equation*}
\frac{\phi^{\text{SNO}}_{\text{CC}}}{\phi^{\text{SNO}}_{\text{NC}}}  =  \snoccncratio. \\
\end{equation*}

\noindent 
The contributions to the systematic uncertainties on the derived fluxes are shown in Table~\ref{errors}.

\begingroup
\begin{table}
\caption{\label{errors}Sources of systematic uncertainties on NC, CC, and ES fluxes.  The total error differs from the individual errors added in quadrature due to correlations.}
\begin{ruledtabular}
\begin{tabular}{llll}
Source      & NC uncert. & CC uncert. & ES uncert. \\ 
         & (\%) & (\%) & (\%) \\  \hline 
PMT energy scale   & $\pm0.6$ & $\pm2.7$ & $\pm3.6$ \\ 
PMT energy resolution   & $\pm0.1$ & $\pm0.1$ & $\pm0.3$ \\ 
PMT radial scaling   & $\pm0.1$ & $\pm2.7$ & $\pm2.7$ \\ 
PMT angular resolution   & $\pm0.0$ & $\pm0.2$ & $\pm2.2$ \\ 
PMT radial energy dep.   & $\pm0.0$ & $\pm0.9$ & $\pm0.9$ \\ 
Background neutrons   & $\pm2.3$ & $\pm0.6$ & $\pm0.7$ \\ 
Neutron capture   & $\pm3.3$ & $\pm0.4$ & $\pm0.5$ \\ 
Cherenkov/AV backgrounds   & $\pm0.0$ & $\pm0.3$ & $\pm0.3$ \\ 
NCD instrumentals & $\pm1.6$ & $\pm0.2$ & $\pm0.2$ \\
NCD energy scale & $\pm0.5$ & $\pm0.1$ & $\pm0.1$ \\
NCD energy resolution & $\pm2.7$ & $\pm0.3$ & $\pm0.3$ \\
NCD alpha systematics & $\pm2.7$ & $\pm0.3$ & $\pm0.4$ \\
PMT data cleaning & $\pm0.0$ & $\pm0.3$ & $\pm0.3$ \\ \hline
Total experimental uncertainty & $\pm6.5$ & $\pm4.0$ & $\pm4.9$ \\ \hline 
Cross section~\cite{crosssection} & $\pm 1.1$  & $\pm 1.2$& $\pm 0.5$ 
\end{tabular}
\end{ruledtabular}
\end{table}
\endgroup

Two independent analysis methods were used as checks of the MCMC method.  Both used maximum likelihood fits but handled the systematics differently.  A comparison of results from these three analysis methods after the blindness conditions had been removed revealed two issues.  A 10\% difference between the NC flux uncertainties was found, and subsequent investigation revealed incorrect input parameters in two methods.  After the inputs were
corrected, the errors agreed and there was no change in the fitted central values.  However, the ES flux from the MCMC fit was 0.5$\sigma$ lower than from the other two analyses.  This difference was found to be from the use of an inappropriate algorithm to fit the peak of the ES posterior distributions.  After a better algorithm was implemented, the ES flux agreed with the results from the other two analyses.  

The ES flux presented here is 2.2$\sigma$ lower than that found by Super-Kamiokande-I~\cite{superk} consistent with a downward statistical fluctuation in the ES signal, as evidenced in the shortfall of signals near $\cos\theta_{\odot}$ = 1 in two isolated energy bins.  The $^8$B spectral shape~\cite{winter} used here differs from that~\cite{ortiz} used in previous SNO results.  The CC, ES and NC flux results in this Letter are in agreement (\textit{p} = 32.8\%~\cite{BLUE}) with the NC flux result of the D$_2$O phase~\cite{ncprl} and with the fluxes from the salt phase~\cite{nsp}.

\begin{figure}
\begin{center}
\includegraphics[width=0.44\textwidth]{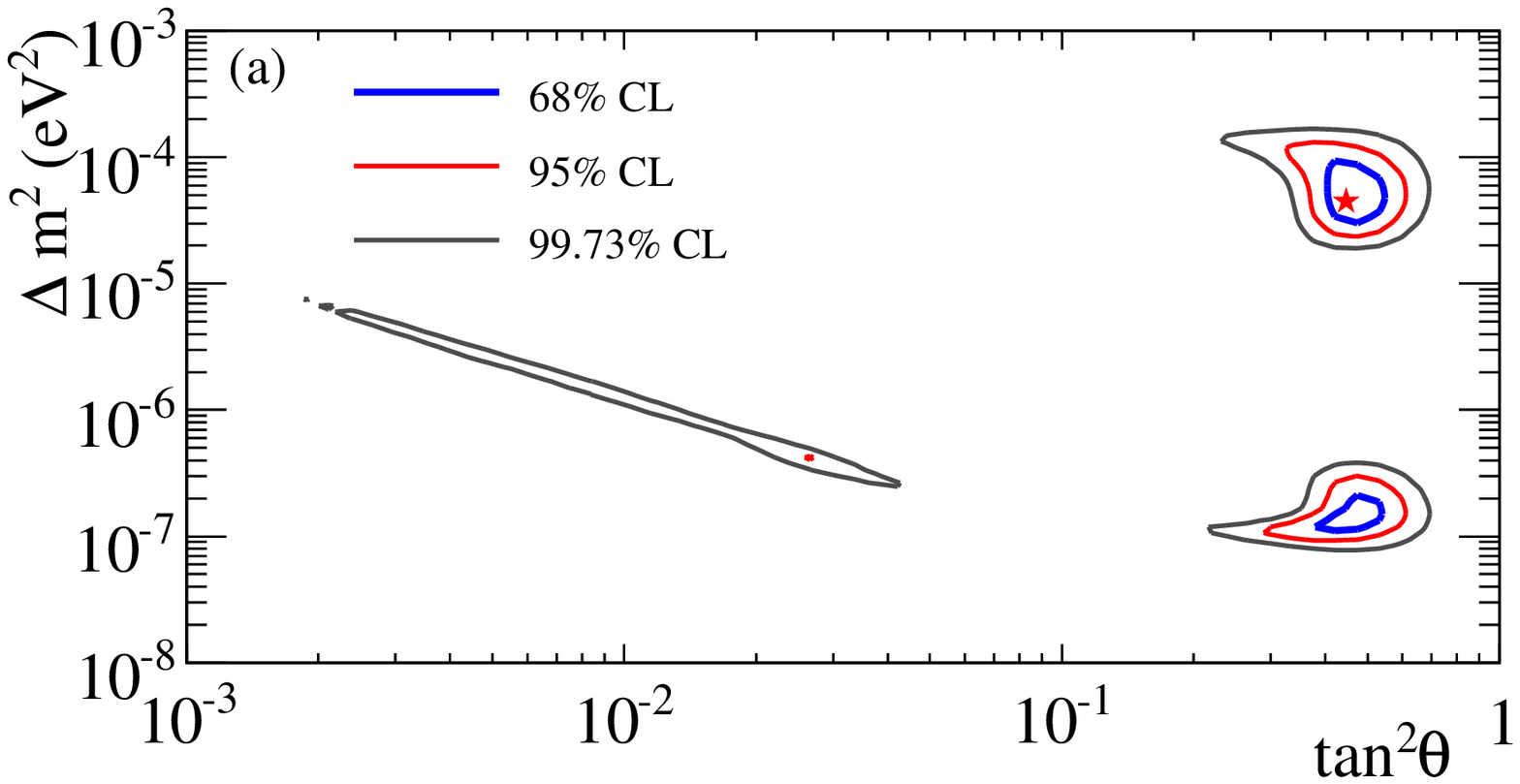}
\includegraphics[width=0.45\textwidth]{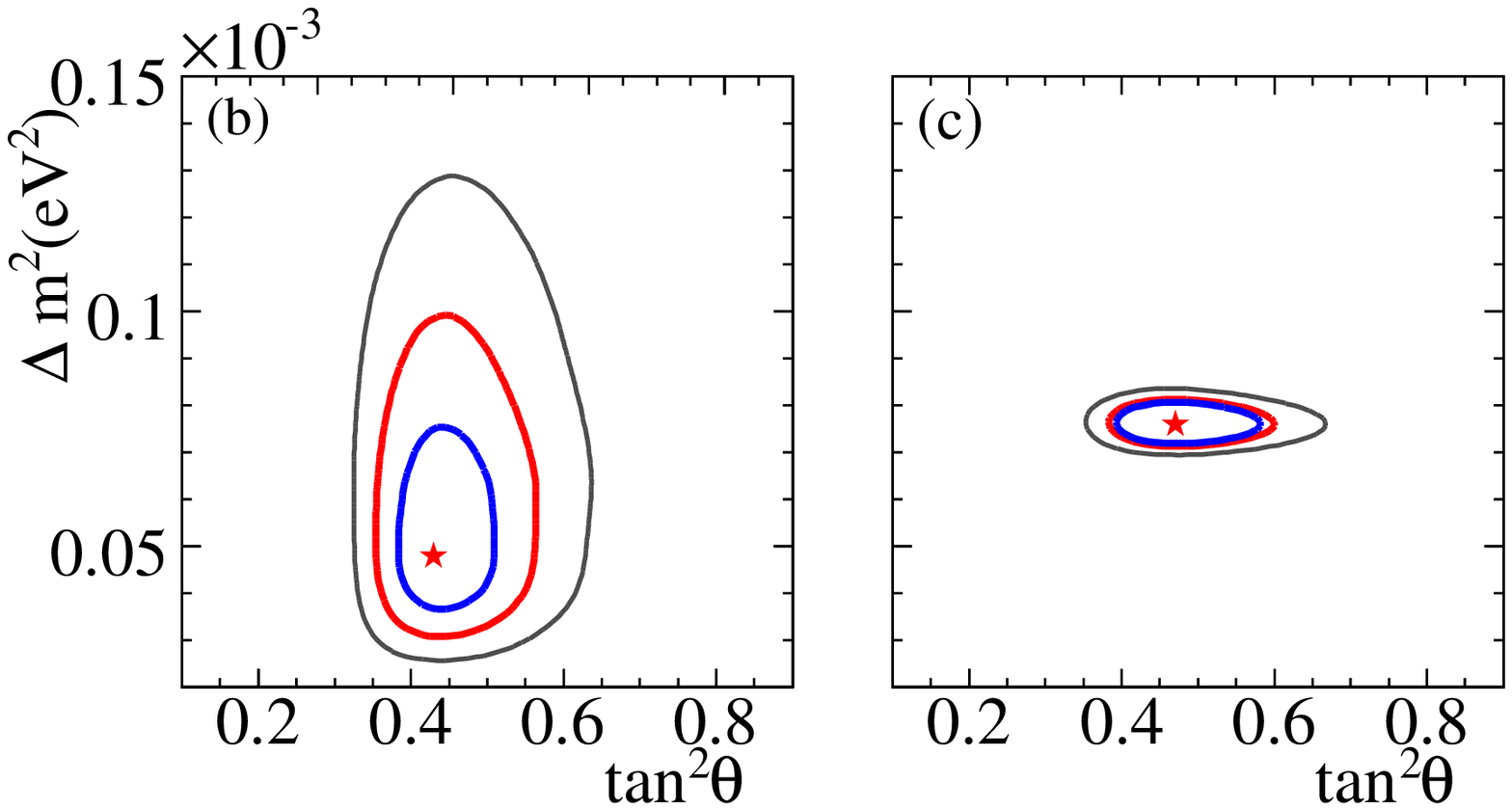}
\caption{\label{globalmsw}Neutrino-oscillation contours. (a) SNO only: D$_2$O \& salt day and night spectra, NCD phase fluxes.  The best-fit point is $\Delta m^2=4.57\times10^{-5}$ eV$^2$, $\tan^{2}\theta=0.447$, $f_{B}=0.900$, with $\chi^{2}$/d.o.f.=73.77/72. (b) Solar Global:  SNO, SK, Cl, Ga, Borexino. The best-fit point is $\Delta m^2=4.90\times10^{-5}$ eV$^2$, $\tan^{2}\theta=0.437$, $f_{B}=0.916$.  (c) Solar Global + KamLAND.  The best-fit point is $\Delta m^2=7.59\times10^{-5}$ eV$^2$, $\tan^{2}\theta=0.468$, $f_{B} = 0.864$.}
\end{center}
\end{figure}

The fluxes presented here, combined with day and night
energy spectra from the pure D$_2$O and salt phases~\cite{dnprl,nsp}, place constraints on
neutrino flavor mixing parameters.  Two-flavor active neutrino
oscillation models are used to predict the CC, NC, and ES rates in SNO~\cite{mnspmsw}.
A combined $\chi^2$ fit to SNO D$_2$O, salt, and NCD-phase data~\cite{snocompanion} yields the allowed regions in $\Delta m^2$ and $\tan^2 \theta$ shown in Fig.~\ref{globalmsw}(a).  
In a global analysis of all solar neutrino data (including Borexino~\cite{borexino} and Super-Kamiokande-I~\cite{superk}) and the 2881 ton-year KamLAND reactor anti-neutrino results~\cite{kam2},
the allowed regions are shown in Fig.~\ref{globalmsw}(b and c).   The best-fit point to the solar global plus KamLAND data yields $\Delta m^{2} = 7.59^{+0.19}_{-0.21}\times10^{-5}$~eV$^2$ and $\theta  = 34.4^{+1.3}_{-1.2}$ degrees, where the errors reflect marginalized 1-~$\sigma$ ranges.
In our analyses, the ratio $f_{B}$ of the total ${}^8$B flux to the SSM~\cite{bs05} value was a free parameter, while the total {\emph{hep}} flux was fixed at $7.93 \times 10^3$~cm$^{-2}$~s$^{-1}$~\cite{bahcall06}.

In summary, we have precisely measured the total flux of active ${}^{8}$B  neutrinos 
from the sun independently from our previous methods. The flux is in agreement with standard solar model calculations.  This Letter presents analysis leading to a reduction in the uncertainty of $\theta$ over our previous results.

This research was supported by:  Canada: NSERC, Industry Canada, NRC,
Northern Ontario Heritage Fund, Vale Inco, AECL, Ontario Power
Generation, HPCVL, CFI, CRC, Westgrid; US: Department of Energy, NERSC PDSF; UK: STFC (formerly PPARC); Portugal: FCT. We thank the SNO technical staff for their strong contributions. 
\vspace{1ex}

\bibliography{ncd_prl}
\end{document}